\documentclass[aps,twocolumn,showpacs,amsmath,floatfix]{revtex4}
\usepackage{graphicx}
\usepackage{dcolumn}
\usepackage{color}
\usepackage{amsmath}
\usepackage{here}
\usepackage{graphicx}
\date{\today}
\begin{document}
\title{Dynamics of two-dimensional coherent structures in nonlocal
nonlinear media}

\author{A. I. Yakimenko$^{1,2}$}%
\author{V. M. Lashkin$^1$}
\author{O. O. Prikhodko$^2$}
\affiliation{$^1$Institute for Nuclear Research, Prospect Nauki
47, Kiev 03680, Ukraine} \affiliation{$^2$Physical Department,
Taras Shevchenko National University, Prospect Glushkova 2, Kiev
03680, Ukraine}

\begin{abstract}
We study stability and dynamics of the single cylindrically
symmetric solitary structures and dipolar solitonic molecules in
spatially nonlocal media. The main properties of the solitons,
vortex solitons, and dipolar solitons are investigated
analytically and numerically. The vortices and higher-order
solitons show the transverse symmetry-breaking azimuthal
instability below some critical power.
 We find the threshold
of the vortex soliton stabilization using the linear stability
analysis and direct numerical simulations. The higher-order
solitons, which have a central peak and one or more surrounding
rings, are also demonstrated to be stabilized in nonlocal
nonlinear media. Using direct numerical simulations, we find a
class of radially asymmetric, dipole-like solitons and show that,
at sufficiently high power, these structures are stable.
\end{abstract}

 \maketitle

\section{Introduction}
The recent experimental observations of spatial solitons in
nonlocal media such as nematic liquid crystals \cite{ContiPRL04},
lead glasses \cite{Segev}, renewed an interest to coherent
structures in spatially nonlocal nonlinear media.
In the spatially nonlocal media the nonlinear response depends on
the wave packet intensity at some extensive spatial domain.
Nonlocality is a key feature of many nonlinear media. It naturally
appears in different physical systems such as plasmas
\cite{LitvakSJPlPhys75,DavydovaFishchukUFZ}, Bose-Einstein
condensates (BEC) \cite{PedrSantos05},  optical media
\cite{KrolikovskiJOptB04}, atomic nuclei \cite{Simenog}, liquid
crystals \cite{ContiPRL04}. 

 An important property of spatially nonlocal nonlinear response is that it
prevents a catastrophic collapse which usually occurs in local
self-focusing media when the power of the two- or
three-dimensional wave packet exceeds some critical value. In
particular, a rigorous proof of absence of collapse during the
wave packet propagation described by the nonlocal nonlinear
Schr\"{o}dinger equation (NLSE) with sufficiently general
symmetric real-valued response kernel has been presented in
\cite{Turitsyn85,KrolikovskiJOptB04}. In the absence of collapse,
the competition between diffraction spreading and nonlinear
self-action leads to formation of the stationary solitary wave
structures - solitons and vortex solitons. Different types of
two-dimensional self-trapped localized wave beams have been
predicted and experimentally observed in various nonlinear media
\cite{KivsharAgrawal}.

Fundamental solitons are the lowest-order localized structures
with a single peak in the intensity distribution and transversely
constant phase. A vortex is the structure with ring-like intensity
distribution, with the dark hole at the center where the phase
dislocation takes place: a phase circulation around the axis of
propagation is equal to $2\pi m$. An integer $m$ is refered to as
topological charge. The important integral of motion associated
with this type of solitary wave is the angular momentum, which can
be expressed through the vortex power $N$ and topological charge
$m$ as follows: $|\textbf{M}|=mN$. Thus, vortices (or spinning
solitons) are the localized structures with nonzero angular
momentum. While the fundamental solitons are robust in a
collapse-free media, the spinning solitons may possess a strong
azimuthal modulational instability \cite{Skryabin98}. As a result,
the vortex decays into several ordinary solitons which fly off
carrying out its energy and angular momentum. Though vortices can
be stabilized in media with competing local nonlinearities
\cite{Quiroga,Berezhiani:01,Mihalache,OurPRE03}, it occurs only in
the extreme regime when the higher-order contribution to the
nonlinearity dominates. The recent investigations
\cite{PRE05,BriedisKrolikowskiVortex} have shown that the
different kinds of nonlocality of the nonlinear response can also
suppress or completely eliminate the symmetry-breaking azimuthal
instability for one-ring and double-ring single-charge ($m=1$)
vortex solitons. The recent experiments \cite{Segev} have
confirmed an existence of stable single-charge vortex solitons in
the the nonlocal media with thermal optical nonlinearity. The
multicharge vortices ($m>1$) are unstable in the media with
thermal nonlocal nonlinearity, as was shown in Ref.~\cite{PRE05}
by linear stability analysis of small azimuthal perturbations and
direct numerical modeling. On the other hand, the robust
propagation of two-charge vortex ($m=2$) solitons has been
observed in the numerical simulations
\cite{BriedisKrolikowskiVortex} for the model based on
Gaussian-type kernel of the nonlocal media response function. We
perform here the linear stability analysis for the model used in
Ref. \cite{BriedisKrolikowskiVortex} and prove the possibility of
stabilization of multicharge vortex solitons in the nonlocal
nonlinear media.

The higher-bound solitons with field nodes (zero-crossing) have
been first discovered in Ref. \cite{Yankauskas} for the local
Kerr-type nonlinear media. The $n$th bound state has a central
bright spot surrounded by $n$ rings of varying size. In the local
nonlinear media the higher-order solitons with zero angular
momentum show the azimuthal instability
\cite{KolokolovSykov,Soto-CrespoPRA91} similar to the instability
of the vortex solitons. The rings which surround the central peak
possess an azimuthal instability. As a result, the higher-bound
structures decay into several fundamental solitons. We reveal here
that nonspinning higher-order solitons can be stabilized in the
nonlocal medium.

Another important feature of nonlocal nonlinear media is the
possibility of existence of composite soliton structures. A
composite soliton structure, or a multi-soliton complex is a
self-localized state which is a nonlinear superposition of several
fundamental solitons \cite{Ankievich, KivsharAsimut}.
Multi-soliton structures in nonlocal media were considered first
in Refs. \cite{Mironov1, Mironov2}, and they have recently
received renewed interest \cite{Nikolov1,Nikolov2,Xu}. In
particular, one-dimensional (1D) nonlocal model suggested in Ref.
\cite{LitvakSJPlPhys75} was studied in Ref. \cite{Xu} and it was
shown that dipole-, triple-, and quadrupole-mode solitons can be
made stable. Very recently \cite{cos} two-dimensional (2D)
rotating dipole structures were considered in the framework of
an approximate variational approach. 
 In this paper, using direct 2D simulations, we find numerically a
class of radially asymmetric two-dimensional dipole-mode soliton
solutions and show that, at sufficiently high input power, these
solutions are stable.

The aim of this paper is to study general properties and to carry
out the stability analysis of single solitons (both fundamental
and higher-bound states), vortex solitons, and composite
dipole-like solitons in the strongly nonlocal media.

The paper is organized as follows. In Sec.~\ref{sec2} we formulate
a model and present basic equations. The stability analysis based
on the variational approach and numerical simulations for single
solitary structures is performed in Sec.~\ref{sec3}, and then, in
Sec.~\ref{sec4}, we consider the multisoliton dipole-like
structures. The conclusions are given in Sec.~\ref{sec5}.

\section{Basic equations}
\label{sec2}
We consider propagation of the electric-field envelope
$\Psi(x,y,z)$ described by the paraxial wave equation:
\begin{equation}\label{eq:NLS}
i\frac{\partial \Psi}{\partial z}+\Delta_\perp\Psi +\Theta \Psi=0,
\end{equation}
where $z$ is the direction of beam propagation, $\Theta$
represents the nonlinear response of the media.

Equation (\ref{eq:NLS}) conserves the following integrals of
motion: (i) number of quanta (``energy" or ``beam power"):
\begin{equation}
 \label{PlNum2D}
 N = \int\left|\Psi \right|^2 d^2 \textbf{r},
\end{equation}
(ii) momentum:
\begin{equation*}
 \label{Momentum2D}
\textbf{I}_\perp = -\frac{\textrm{i}}{2} \int
 \left( \Psi^* \nabla_\perp \Psi - \Psi \nabla_\perp \Psi^* \right)
 d^2 \textbf{r},
\end{equation*}
(iii) angular momentum:
\begin{equation*}
\textbf{M}= -\frac{\textrm{i}}{2} \int \left[\textbf{r}
\times\left( \Psi^*\nabla_\perp \Psi- \Psi\nabla_\perp
\Psi^*\right)\right] d^2 \textbf{r},
\end{equation*}
(iv) Hamiltonian:
\begin{equation*}  \label{Hamilt2D}
 H = \int\left\{| \nabla_\perp \Psi|^2-\frac12\Theta |\Psi|^2
 \right\}d^2\textbf{r}.
\end{equation*}

The nonlocal nonlinear media response function can be taken as
follows:
\begin{equation}  \label{eq:Theta}
\Theta(\vec r)=\int{R(|\vec{r}-\vec{r_1}|)
|\Psi(\vec{r_1})|}^2d^2\textbf{r}_1.
\end{equation}

The shape of the kernel $R(r)$ is determined by the type of the
nonlocal interaction in media and can be rather complicated
\cite{PRE05,PedrSantos05}. However, there are general properties
valid for all nonlocal media response functions. The nonlinear
term tends to the local Kerr-type nonlinearity:
$\Theta\to|\Psi|^2$ when the spatial scale of the wave packet
intensity distribution $|\Psi|^2$  is much wider than the
effective width of the potential $R(r)$. In the opposite case of
the strongly nonlocal regime, the response function can be
estimated as follows: $\Theta(r)=N\left\{R(0)+\frac12\Delta
R(0)r^2\right\}$. In the latter case, the highly-nonlocal NLSE
(\ref{eq:NLS}) with a sufficiently regular kernel $R(r)$ is
mathematically identical to the linear Schr\"{o}dinger equation
with harmonic oscillator potential, as was pointed out in Ref.
\cite{SnyderMitchellScience}.

We consider in this paper the nonlocal response function kernel
modeled by the Gaussian shape kernel:
\begin{equation}
\label{kernel}
 R(|\vec{r}-\vec{r_1}|)=\frac{\alpha^2}{\pi}
e^{-\alpha^2\left|\vec{r}-\vec{r_1}\right|^2},
\end{equation}
where $\alpha$ is the nonlocality parameter. Keeping the main
features of nonlocal media this model  allows  a very accurate and
simple analytical treatment.
\section{Single solitary structures}
\label{sec3}
In this section, we study single solitons (both fundamental and
higher-bound states) and vortex solitons. We look for the
stationary solutions of the Eq. (\ref{eq:NLS}) in the form:
\begin{equation}\label{eq:psistationary}
\Psi(x,y,z)=\psi(r) e^{im\varphi+i\Lambda z},
\end{equation}
where $\varphi$ and $r=\sqrt{x^2+y^2}$ are the azimuthal angle and
the radial coordinate, respectively, and $\Lambda$ is the beam
propagation constant. Such solutions describe either the soliton,
when $m=0$, or the vortex soliton with the topological charge $m$,
when $m \neq 0$. The function $\psi(r)$ obeys the
integro-differential equation:
\begin{equation}\label{eq:psi_Vs_r}
-\Lambda\psi+\Delta_r^{(m)}\psi+\theta\,\psi=0.
\end{equation}
The boundary conditions for the localized solutions are:
$\psi(\infty)=0$ and $(d\psi/dr)_{r=0}=0$ for solitons, $\psi(0)
=0$ for
vortices. 
For the stationary solution (\ref{eq:psistationary}) it is easy to
rewrite the response function in the form:
\begin{equation}\label{eq:thetaSteady}
 \theta(r)=2\alpha^2\int_0^{+\infty}e^{-\alpha^2(r-r_1)^2}\mathcal{I}_0(2\alpha^2 r r_1)|\psi(r_1)|^2 r_1d
 r_1,
\end{equation}
where $\mathcal{I}_\nu(x)=e^{-x}I_\nu(x)$ is the exponentially
scaled modified Bessel function.


 In the next subsection we
start our considerations on the single cylindrically symmetric
solitary structures with the analytical analysis based on the
variational approach.

\subsection{Variational approach}
As known \cite{SnyderMitchellScience}, the nonlocal NLSE turns to
the linear Schr\"{o}dinger equation with harmonic oscillator
potential in the highly-nonlocal limit, when the spatial scale of
the response function is much wider than the wave packet
localization region. Since the Laguerre-Gauss modes are the exact
eigenstates for the two-dimensional linear oscillator, the
variational method with the trial function of the form
\begin{equation}\label{ansatz}
\Psi(x,y,z)=h(z)\xi^mL_n^{(m)}\left(\xi^2
\right)e^{-\frac12\xi^2\left\{1+i\tilde
b(z)\right\}+im\varphi+i\Phi(z)},
\end{equation}
 is expected to give an accurate
description of all eigenmodes, especially in the highly nonlocal
regime. Here $L_n^{(m)}(x)$ is  the generalized Laguerre
polynomial, $n$ is the number of nodes of the radial profile, and
$m$ is the topological charge, $\xi=r/a(z)$, where $a(z)$ is the
first variational parameter that characterizes a radius of
solitary structure: $a^2=\langle r^2\rangle(2n+m+1)^{-1}$, where
$\langle r^2\rangle=N^{-1}\langle\Psi| r^2|\Psi\rangle$ is the
mean-square radius. The second variational parameter $\tilde b(z)$
is the phase curvature. The amplitude $h(z)$ can be readily found
from the relation $h(z)a(z)=\sqrt{\frac{n!N}{\pi (n+m)!}},$ which
is obtained from normalization condition (\ref{PlNum2D}).

We start our considerations with the lower-order node-less states
($n=0$). The nonlinear response function in the framework of
variational approach is given by expression:
$$\Theta(r)=h^2 q e^{-q\xi^2}m! (1-q)^mL_m\left(q^2\xi^2/(q-1)\right),$$
where $q=\alpha^2a^2/(1+\alpha^2a^2)$, $L_m(x)=L_m^{(0)}(x)$ is
Laguerre polynomial of $m$-th order. Note, if $a^2\alpha^2\gg 1$,
then $q\to 1$, and one obtains $\Theta\to |\Psi|^2$, as it should
be in the local limit.

In accordance with the variational procedure, we construct the
Lagrangian density
$$
\ell=\frac{i}{2}\left(\Psi\frac{\partial\Psi^*}{\partial z}-
\Psi^*\frac{\partial\Psi}{\partial z}\right)+
\left|\nabla_\perp\Psi\right|^2-\frac12\Theta|\Psi|^2
$$
and the Lagrangian
$$\mathcal{L}=\int \ell~ d^2 \textbf{r}=N\dot{\Phi}+\frac{(m+1)}{2}N\left[b \dot{a}-\dot{b}a\right]+H,$$
where $b=\tilde b/a$, $H$ is the Hamiltonian:
\begin{equation*}\label{eq:hamiltvariat}
H=N\left\{(m+1)\left(1/a^2+b^2\right)-\frac{N h_m(a)}{2\pi
a^2}\right\},
\end{equation*}
 where
$h_m(a)=q\left(1-q\right)^m\left(1+q\right)^{-m-1}P_m\left(\frac{1+q^2}{1-q^2}\right)$,
$P_m(x)$ is the Legendre polynomial of the $m$-th order.

The first two dynamical equations can be written in the canonical
form:
\begin{equation}\label{eq:dynamvariat}
\frac{N(m+1)}{2}\frac{da}{dz}=-\frac{\partial H}{\partial b},~~~
\frac{N(m+1)}{2}\frac{d b}{dz}=\frac{\partial H} {\partial a}
\end{equation}
 and
the third one is the following: $dN/dz=0$ which means that the
number of quanta is the integral of motion. The soliton or vortex
soliton corresponds to the stationary point of the Hamiltonian:
$\partial H/\partial b=0$, $\partial H/\partial a=0$. The first
condition  yields $b_0=0$. From the second equation one can find
the width $a_0$ as the function of the number of quanta $N$. It is
easy to verify that vortex or soliton can exist only above the
threshold value for number of quanta:
$N>N_{cr}=4^{m+1}\pi(m+1)(m!)^2/(2m)!$. Using the similar
procedure we have considered the nonspining ($m=0$) higher-bound
states with one ($n=1$) and two nodes ($n=2$). The thresholds for
an existence of the higher bound states are as follows:
$N_{cr}=24\pi$ for $n=1$ and $N_{cr}=640\pi/11$ for $n=2$.

The described variational procedure provides the possibility for
analysis of the stationary radially-symmetric coherent structures.
The results of the variational analysis  are given in Fig.
\ref{fig:n0} (a) for fundamental solitons and vortex solitons, and
in Fig. \ref{fig:m0n1n2} for higher-order solitons.

Moreover, using the set (\ref{eq:dynamvariat}) it is possible to
study the radially-symmetric dynamics of the localized wavepackets
propagating in $z$-direction. Let us investigate, for example, the
evolution of a slightly perturbed stationary soliton solution. It
can easily be shown that the small deviation of the soliton width
$\delta=a-a_0$ from the stationary value $a_0$ obeys the equation
$\ddot{\delta}+\omega^2\delta=0$, where
$\omega^2=8N^{-1}(\partial^2 H/\partial a^2)_{a_0,b_0}.$
Therefore, the soliton being radially perturbed exhibits the
oscillations with the frequency
$\omega=8\alpha^2\pi^{-1/2}[1-\sqrt{N_{cr}/N}]^{3/2},$ where
$N_{cr}=4\pi$ is the threshold for soliton existence.

However, one cannot study the stability of stationary solutions
with respect to symmetry-breaking azimuthal perturbations in the
framework of a variational approach with a radially symmetric
trial function. Stability conditions of steady-state solutions,
regarding small general 2D perturbations, will be obtained by a
linear stability analysis in the next subsection.
\subsection{Numerical modeling}
\begin{figure*}
\includegraphics[width=6.8in]{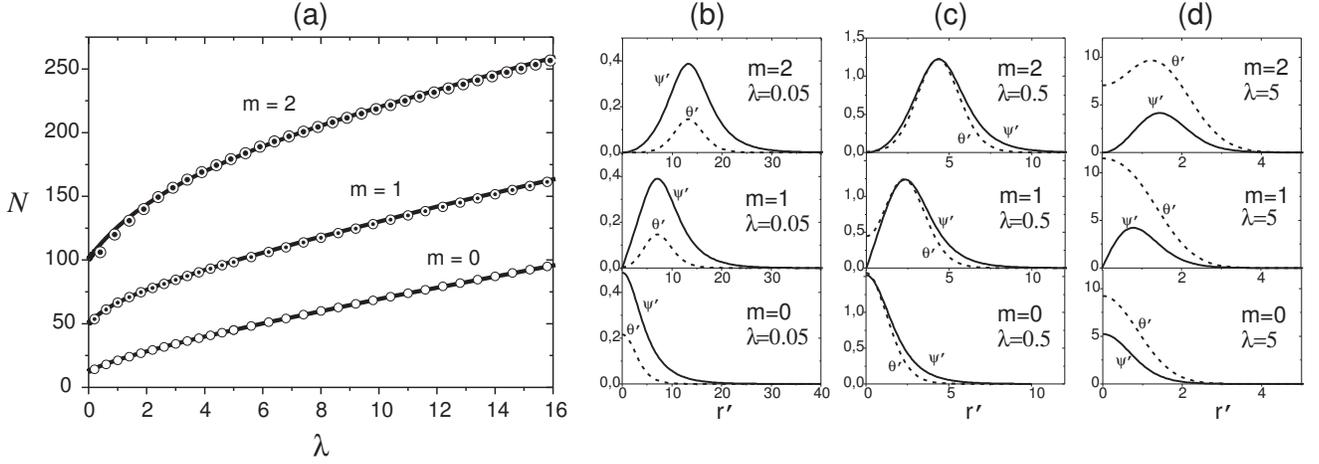}
\caption{(a) Number of quanta $N$ vs parameter
$\lambda=\Lambda/\alpha^2$ for  solitons  ($n=0,m=0$) and vortex
solitons  ($n=0,m=1, 2$) (solid curves for variational and circles
for numerical results). Numerically found profiles for (b)
$\lambda=0.05$; (c) $\lambda=0.5$; (d) $\lambda=5$. Solid curves
for $\psi'(r')=\psi/\alpha$ dashed curves for
$\theta'(r')=\theta/\alpha^2$; the scaled coordinates $r'=\alpha
r$ are used.} \label{fig:n0}
\end{figure*}

The boundary problem (\ref{eq:psi_Vs_r}) is  equivalent to the
integral equation:
\begin{equation}\label{eq:E_rintegralEq}
\psi(r)=\int_0^{+\infty}\theta(\eta)
\psi(\eta)\,G_m(\eta,r;\sqrt{\Lambda}) \eta\,d\eta,
\end{equation}
where $\theta$ is given by Eq. (\ref{eq:thetaSteady}),
\begin{equation}
\label{eq:Green'sfunction} G_m(\xi_1,\xi_2; a)=\left\{
  \begin{array}{lc}
    K_m(a\xi_2)I_m(a\xi_1), & 0\le \xi_1<\xi_2, \\
    I_m(a\xi_2)K_m(a\xi_1), & \xi_2<\xi_1<+\infty,
  \end{array}
\right.
\end{equation}
where $I_m$ and $K_m$ are the modified Bessel functions of the
first and second kind, respectively.
 We have solved the nonlinear integral equation
(\ref{eq:E_rintegralEq}) using stabilized iterative method
\cite{Petviashvili86}. For numerical modeling it is useful to
perform the rescaling transformation of the form: $r'=\alpha r$,
$\psi'=\psi/\alpha$, $\theta'=\theta/\alpha^2$, $z'=z\alpha^2$.
Such transformation reduces the number of parameters to one
dimensionless parameter $\lambda=\Lambda/\alpha^2$. Figure
\ref{fig:n0} (b)-(d) shows several examples of the numerical
solution of
 the (\ref{eq:psi_Vs_r}) at different values of the parameter
 $\lambda$. Note, that the nonlocal limit $\alpha^2\ll\Lambda$
 corresponds to the large values of the parameter $\lambda$ and,
  as seen from
 Fig. \ref{fig:n0} (a), to the large values of the beam power
 $N$.

\begin{figure*}
\includegraphics[width=6.8in]{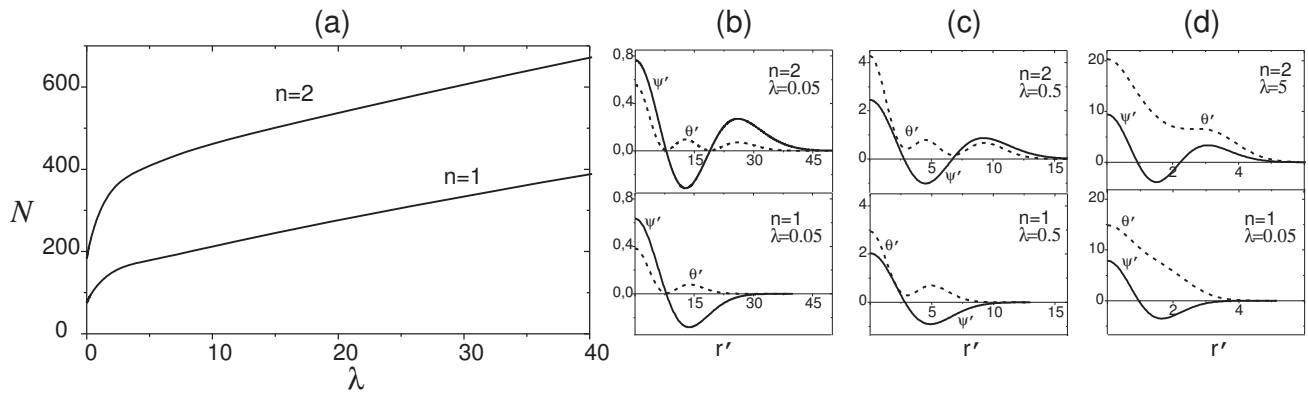}
\caption{(a) Number of quanta $N$ vs parameter
$\lambda=\Lambda/\alpha^2$ for higher-order solitons ($m=0$,
$n=1,2$). Radial profiles for (b) $\lambda=0.05$; (c)
$\lambda=0.5$; (d) $\lambda=5$. Solid curves for $\psi'$ dashed
curves for $\theta'$. }\label{fig:m0n1n2}
\end{figure*}

\begin{figure}
\includegraphics[width=3.4in]{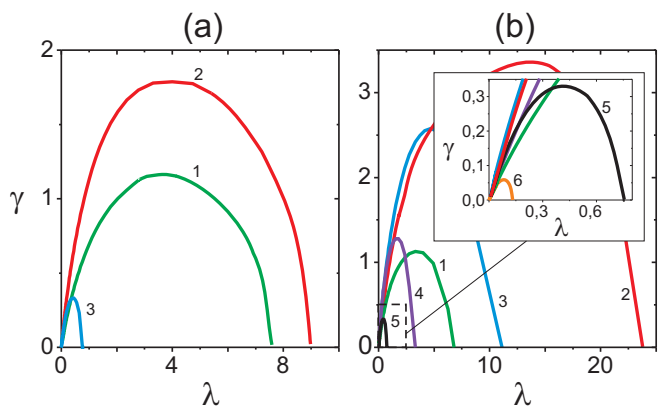}
\caption{(Color online) The scaled growth rate
$\gamma=\textrm{Im}\omega/\alpha^2$ of linear perturbation modes
vs the parameter $\lambda=\Lambda/\alpha^2$ for the vortices with
(a) $m=1$ and (b) $m=2$, the inset depicts the growth rates of the
high-$L$ modes in more details. The numbers near the curves stand
for the azimuthal mode numbers $L$.} \label{fig:GrRates}
\end{figure}

 Let us investigate the stability of the steady-states with respect to small azimuthal perturbations.
 Expanding the nonstationary solution in
 vicinity of a steady-state:
 \begin{eqnarray}
 \label{linearPerturb}\nonumber
 \Psi(r,z)=\left\{\psi(r)+\delta\psi_+ +\delta\psi^*_-\right\}e^{i\Lambda
 z+im\varphi},
 \end{eqnarray}
where $\delta\psi_\pm(r,\varphi,z)=\varepsilon_\pm(r)e^{i\omega
z+i
 L\varphi}$,  and linearizing the dynamical Eq. (\ref{eq:NLS}) one
can obtain the eigenvalue problem for $\omega$ of the form:
 \begin{equation}
 \label{eq:EigenProblem}
  \begin{pmatrix}
    \hat Q_{m+L}+\hat g_L & \hat g_L \\
-\hat g_L     & -\hat Q_{m-L}-\hat g_L
  \end{pmatrix}
\vec{\varepsilon}=\omega\vec{\varepsilon},
 \end{equation}
where $\vec\varepsilon=(\varepsilon_+,\varepsilon_-)$, and
$|\textrm{Im}\omega|$ determines the growth rate of an unstable
mode,
$$\hat Q_{m\pm L}\varepsilon_\pm=\left\{-\Lambda+
\Delta_r^{(m\pm L)}+\theta(r)\right\}\varepsilon_\pm,$$ $$\hat
g_L\varepsilon_\pm=2\alpha^2~\psi(r) \int_0^{\infty}
\psi(\xi)\chi(r,\xi)\varepsilon_{\pm}(\xi) d\xi,
$$
$\chi(r,\xi)=\xi e^{-\alpha^2(r-\xi)^2}\mathcal{I}_L(2\alpha^2 r
\xi)$.   The unperturbed radial profile $\psi(r)$ is taken to be
real without loss of generality. We have used the Hankel spectral
transformation and reduced the integro-differential eigenvalue
problem (\ref{eq:EigenProblem}) to the linear algebraic one.
Figure \ref{fig:GrRates} (a) shows the maximum growth rate for
one-charge ($m=1$) vortex solitons. It is seen that only modes
having $L=1,2,3$ can be unstable. Modulational instability is
strongly suppressed in a highly-nonlocal regime: the growth rates
vanish at some finite values of parameter $\lambda$. The mode with
azimuthal number $L=2$ corresponds to the largest growth rates
with widest instability region: all growth rates are equal to zero
at $\lambda>\lambda_{cr}\approx 9.1$. Similar analysis has been
performed for multi-charge vortices (with $m=2,...5$). Figure
\ref{fig:GrRates} (b) depicts the growth rates for two-charge
($m=2$) vortex solitons. Note that the modulation instability is
eliminated for $\lambda>\lambda_{th}\approx 23.8$. Thus, in the
media with the nonlocal response of the form (\ref{kernel}) the
multi-charge vortices can be stabilized. Note that vortex solitons
with $m>1$ are found \cite{PRE05} to be unstable in the media with
thermal nonlocal nonlinearity. Therefore, the dynamical properties
of the vortex structures can be significantly affected by specific
type
of the nonlocal interaction.

\begin{figure}
\includegraphics[width=3.4in]{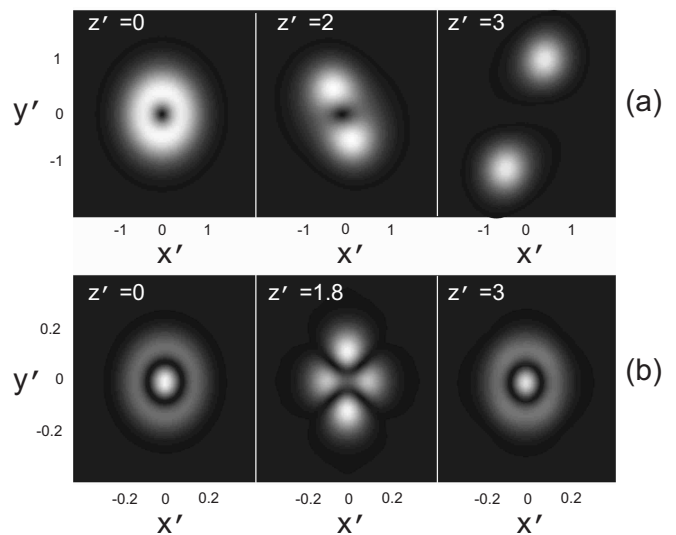}
\caption{Snapshots of $|\Psi'(x',y')|$ for different $z'$. (a)
Example of unstable evolution of single-charge ($m=1$, $n=0$)
vortex soliton with $\lambda=4$. (b) Dynamics with revival of the
non-spinning ($m=0$) one-node ($n=1$) soliton at $\lambda=35$.}
\label{fig:dynam}
\end{figure}

The results of linear analysis have been confirmed by extensive
series of numerical simulations of dynamics of perturbed
stationary solutions. We have performed the split-step Fourier
transform method to solve the dynamical Eq. (\ref{eq:NLS}) with
the response function Eq. (\ref{eq:Theta}). The nonlocal nonlinear
term has been calculated in the spectral domain, since it has the
form of the convolution of the intensity distribution
$|\Psi(\vec{r})|^2$ with the function $R(|\vec{r}|)$. We have used
the numerically found stationary vortex solitons and variational
profiles for the higher-order solitons as the initial conditions.
Different kinds of the perturbations such as random noise,
radially-symmetric and azimuthaly periodical perturbations have
been applied in the numerical experiments. The conclusions of the
linear stability analysis are found to be in a good agreement with
our simulations: the vortex solitons become stable above some
critical power which is close to the one predicted by linear
stability analysis.

Figure \ref{fig:dynam} (a) illustrates the unstable evolution of
the single-charge vortex soliton. In the stable region, when
$\lambda>\lambda_{cr}$, vortices survive even being strongly
perturbed. The mean-square radius and intensity of the vortex
oscillate, but the vortex ring shows the robust propagation over
vast distances (thousands of diffraction lengths) for the hundreds
of the effective periods of the oscillation. The higher-order
solitons exhibit similar decay in weakly nonlocal regime. However,
when beam power increases, an interesting dynamics with revivals
has been observed [see Fig. \ref{fig:dynam} (b)]. Initially field
envelope decays into several filaments, but then it recurs at
larger propagation distances. With further increasing of the
power, a higher-order soliton occurs to be stabilized: it shows
the robust
 propagation without breakup during the hundreds of effective
oscillation periods.

\section{Stable bisoliton molecules}
\label{sec4}

\begin{figure}
\includegraphics[width=3.4in]{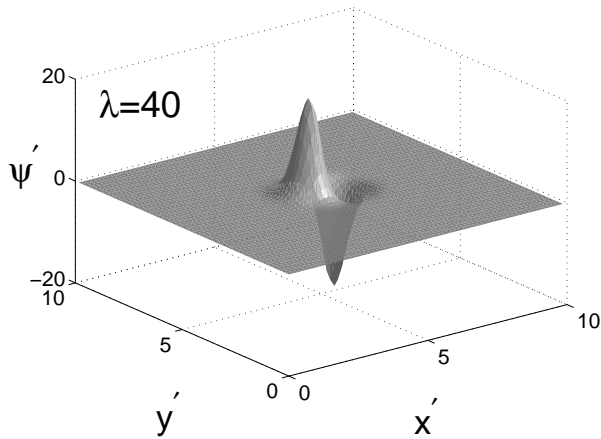}
\caption{\label{fig1} The dipole solution $\psi'=\psi/\alpha$ for
$\lambda\equiv \Lambda/\alpha^{2}=40$; the scaled coordinates
$x'=\alpha x$ and $y'=\alpha y$ are used.}
\end{figure}

In this section, we present and study localized asymmetric
dipole-like solutions. We look for stationary solutions of Eq.
(\ref{eq:NLS}) in the form $\Psi(x,y,z)=\psi(x,y)\exp(i\Lambda
z)$, so that $\psi(x,y)$ obeys the equation
\begin{equation}
\label{stat}
 -\Lambda\psi+\Delta_{\bot}\psi+\theta\,\psi=0,
\end{equation}
where
\begin{equation}
\label{teta} \theta=\frac{\alpha^{2}}{\pi}\int
e^{-\alpha^{2}[(x-x_{1})^{2}+
(y-y_{1})^{2}]}\psi^{2}(x_{1},y_{1})d^2\mathbf{r}_{1}.
\end{equation}
 and we do not assume the radial symmetry of $\psi(x,y)$.
Unlike the NLSE with local cubic nonlinearity, equation
(\ref{stat}) with the nonlocal nonlinear media response $\theta$
given by Eq. (\ref{teta}) has two characteristic transverse
scales: the internal scale $\Lambda^{1/2}$ (in the NLS this scale
determines the characteristic size of the soliton) and the
"external" scale $\alpha$ which is the measure of nonlocality.
Under this, the characteristic size of the self-consistent
potential well in Eq. (\ref{eq:NLS}) can significantly differ from
$\Lambda^{-1/2}$ and, thus, the existence of composite soliton
structures becomes possible. In this paper we restrict ourselves
to dipole-like localized solutions. Results concerning the
tripolar and higher radially asymmetric soliton modes will be
published elsewhere.

As above, we use the scaled variables $\psi'$, $z'$, $x'=\alpha
x$, and $y'=\alpha y$. Imposing periodic boundary conditions on
Cartesian grid and choosing an appropriate initial guess, one can
find numerically dipole-type localized solutions by using the
relaxation technique similar to one described in Ref.
\cite{Petviashvili86}. An example of such dipole solution is
presented in Fig.~\ref{fig1}. The dipole consists of two
out-of-phase monopoles. The characteristic width of the monopoles
in the dipole and the "distance" between them decrease with
increasing the parameter $\lambda=\Lambda/\alpha^{2}$.

\begin{figure}
\includegraphics[width=4.2cm]{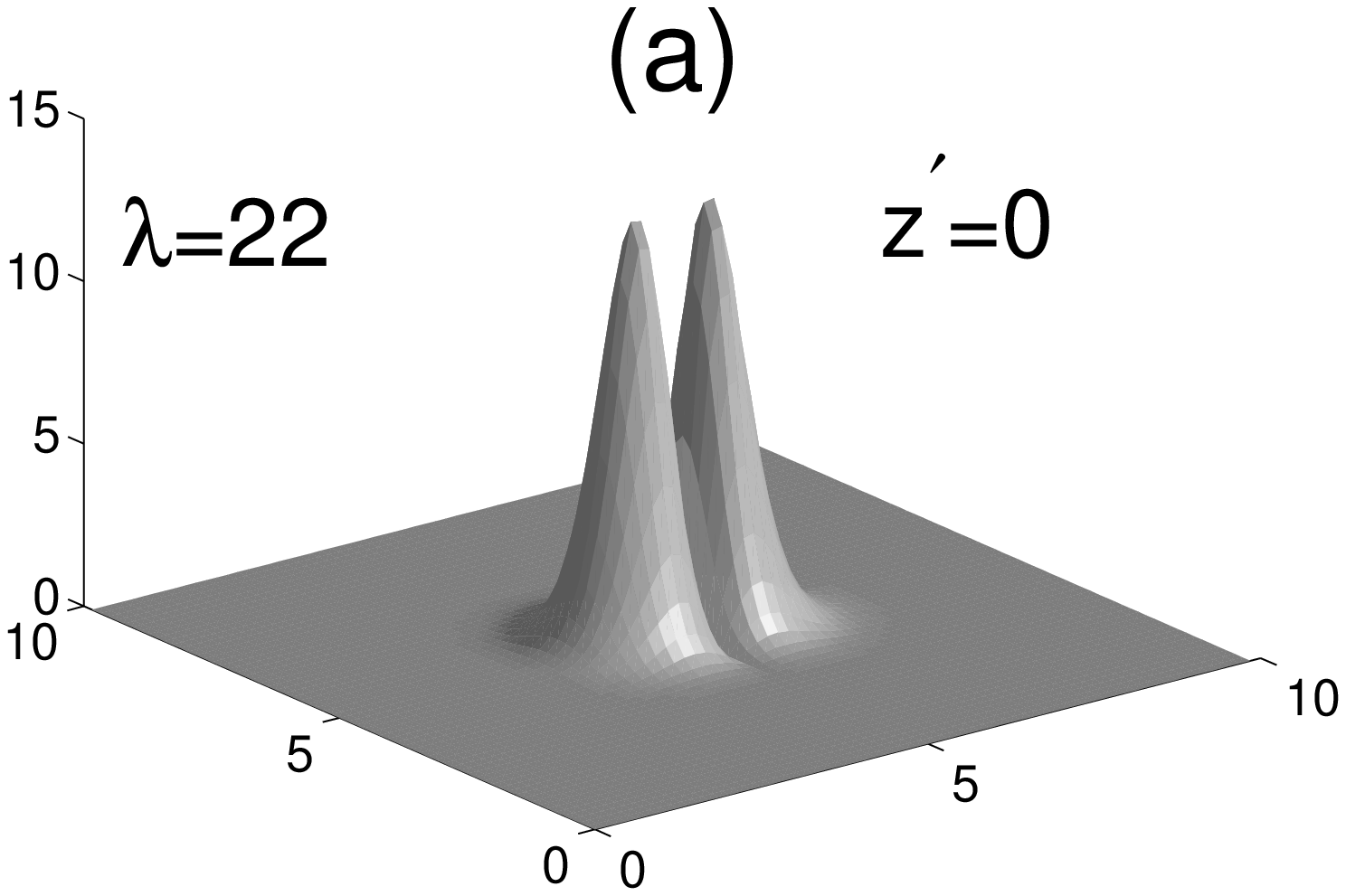}
\includegraphics[width=4.2cm]{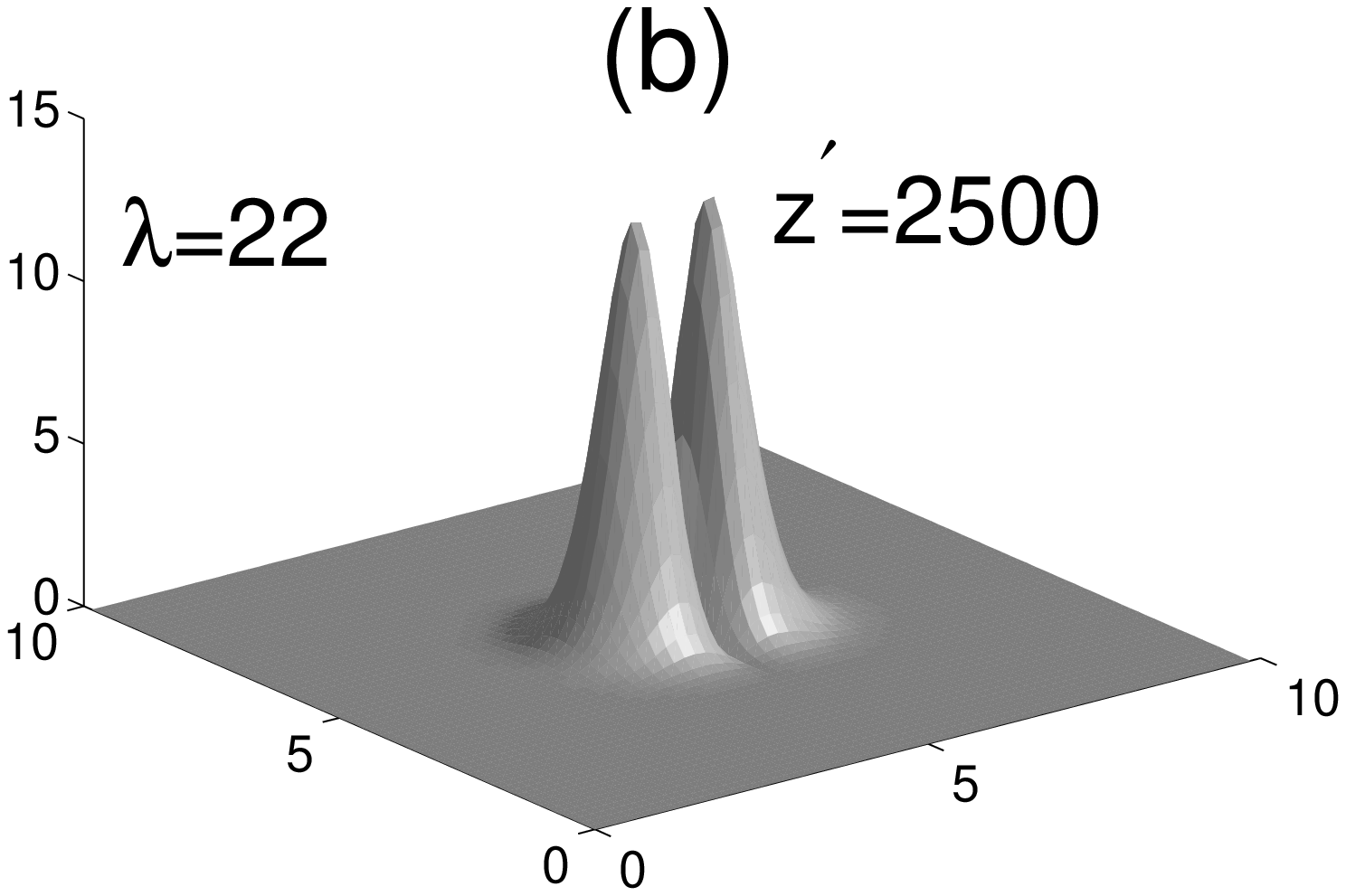}
\includegraphics[width=4.2cm]{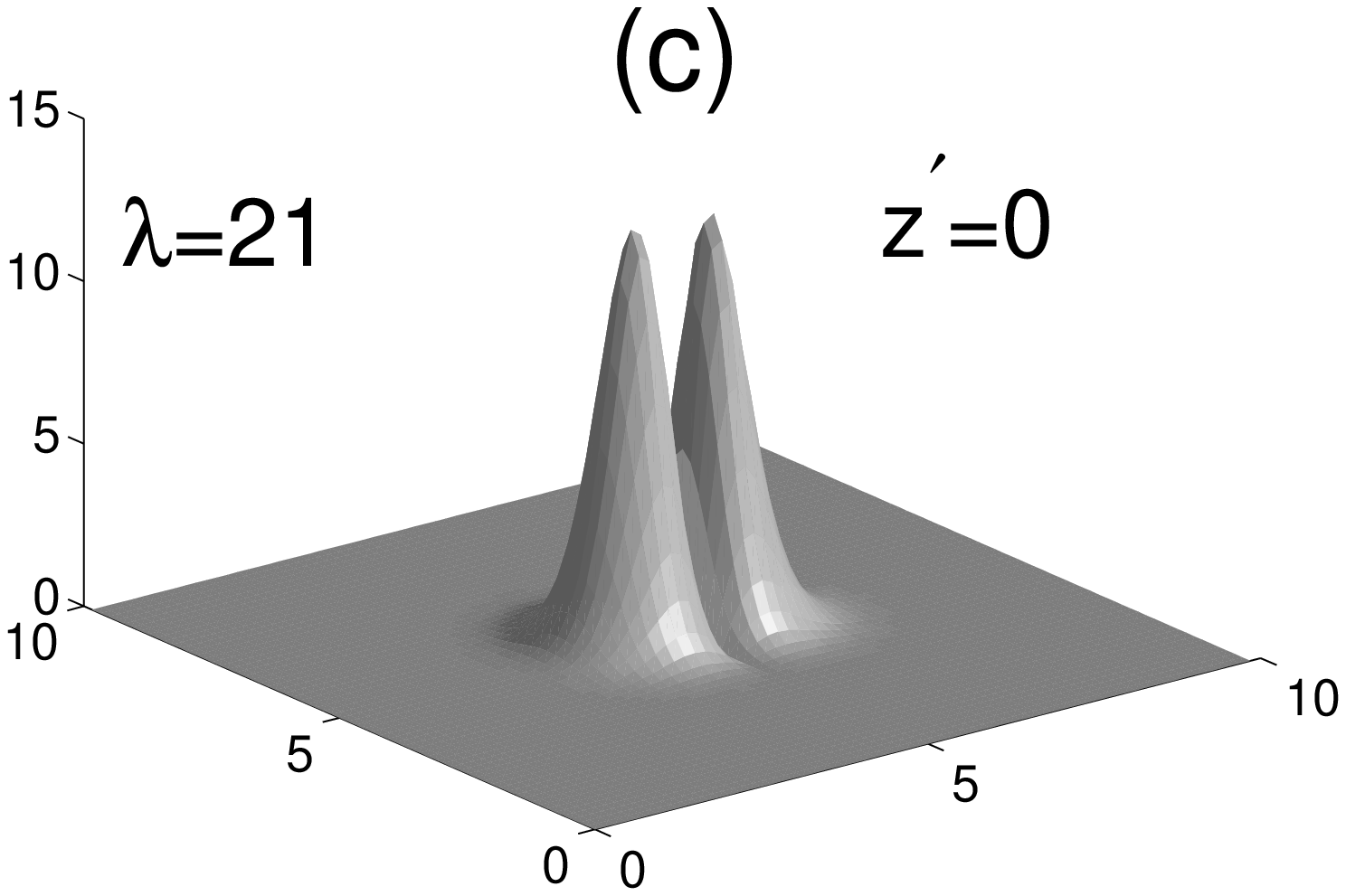}
\includegraphics[width=4.2cm]{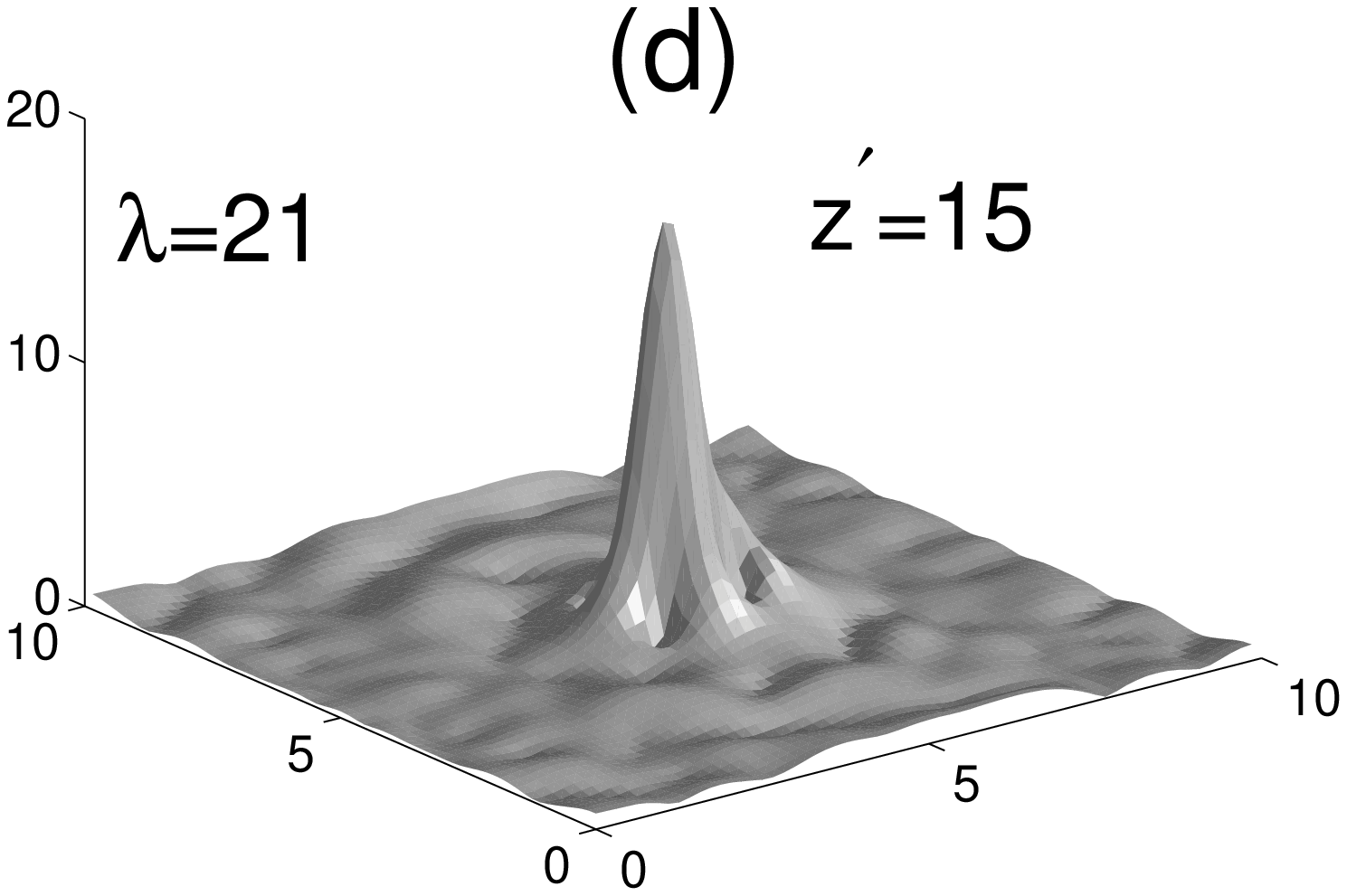}
\includegraphics[width=4.2cm]{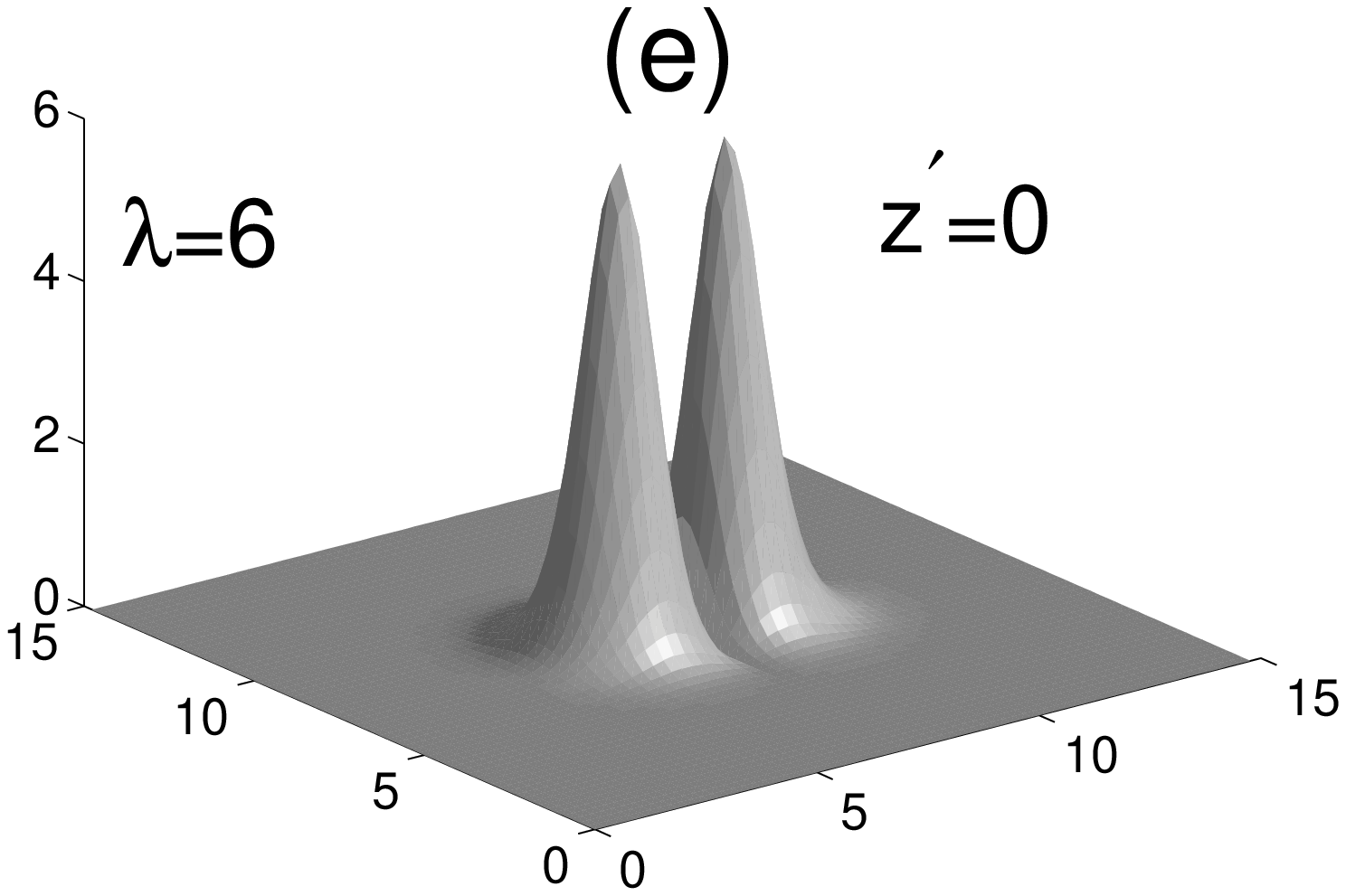}
\includegraphics[width=4.2cm]{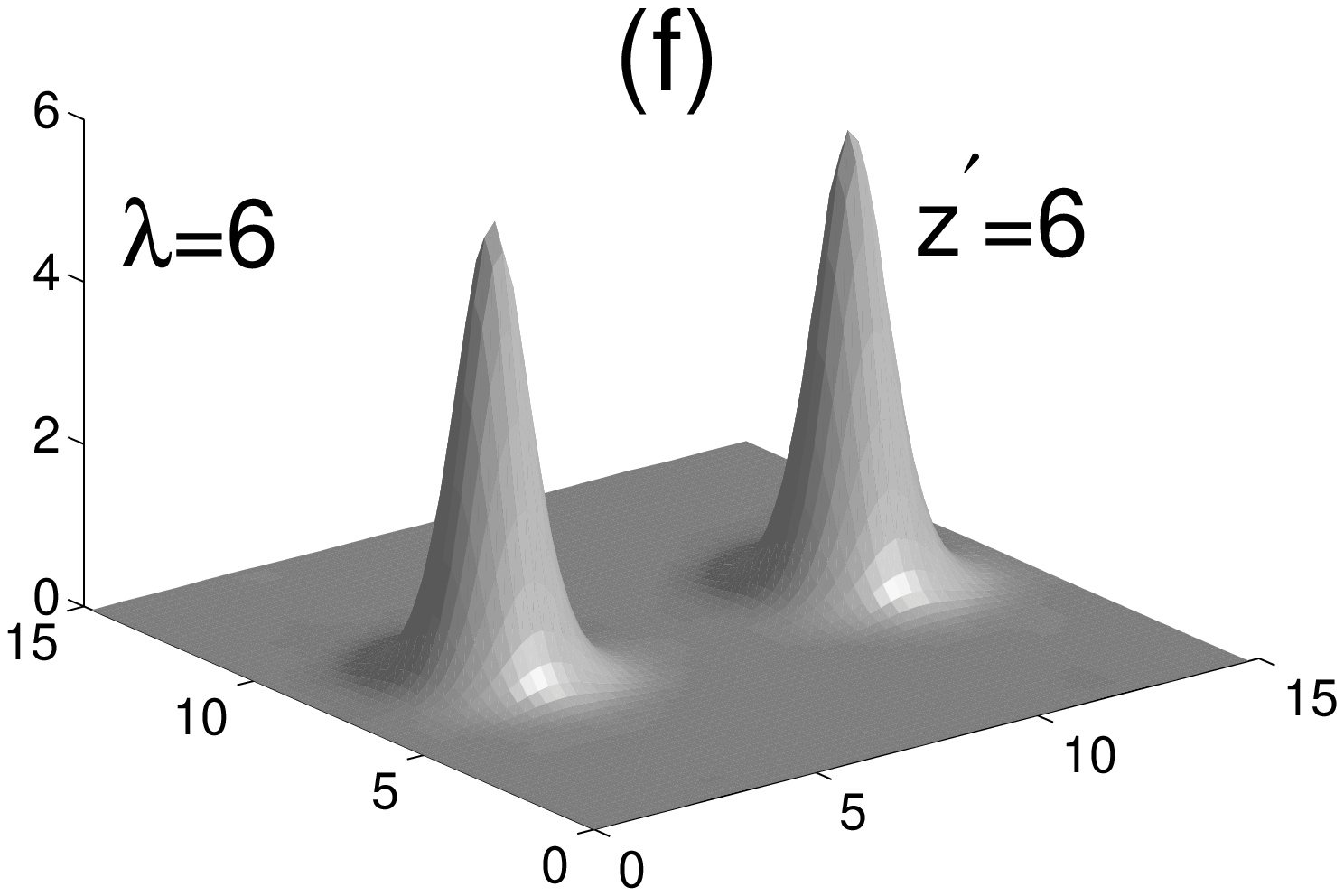}
\caption{\label{fig2} Snapshots of $|\psi'(x',y')|$ for dipole
propagation in the presence of small random perturbation for
different $\lambda=\Lambda/\alpha^{2}$, shown in the scaled
$(x'=\alpha x,y'=\alpha y)$ plane, at different moments $z$: (a),
(b) - stable propagation; (c), (d) - destruction of the original
dipole; (e), (f) - decay into two solitons.}
\end{figure}

\begin{figure}
\includegraphics[width=3.4in,height=4cm]{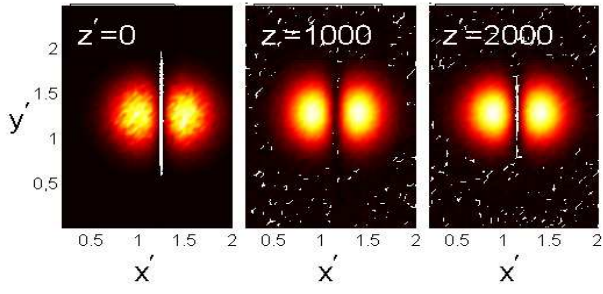}
\caption{\label{fig0} (Color online) Evolution of the dipole
solution with $\lambda=400$ in the presence of strong initial
random perturbation; the scaled variables $x'=\alpha x$,$y'=\alpha
y$ and $z'=\alpha^{2}z$ are used.}
\end{figure}

\begin{figure}
\includegraphics[width=3.4in]{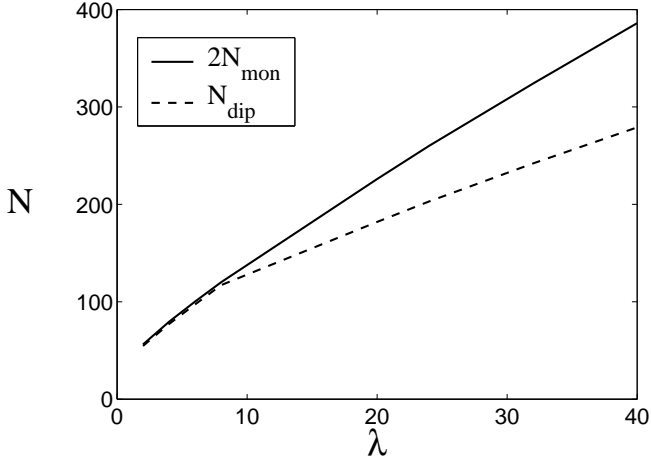}
\caption{\label{fig3} The doubled energy of the soliton (monopole)
$2N_{mon}$ (solid line) and the energy of the dipole $N_{dip}$
(dashed line) v.~s. the parameter $\lambda=\Lambda/\alpha^{2}$.}
\end{figure}

We next addressed the stability of these dipole solutions and
study the evolution (propagation) of the dipoles in the presence
of small initial perturbations. We have undertaken extensive
numerical modeling of Eqs. (\ref{eq:NLS}), (\ref{eq:Theta}) and
(\ref{kernel}) initialized with our computed dipole-type solutions
with added gaussian noise. Spatial discretization was based on the
pseudospectral method 
and "temporal" $z$-discretization included the split-step scheme.
The numerical simulations clearly show that the dipoles with
$\lambda
> \lambda_{th}$, where $\lambda_{th}\approx 21$ is the threshold value,
are stable with respect to small initial noisy perturbations up to
the maximum propagation distances used (of the order of
$z'=3000$). The stable propagation of the dipole is illustrated in
Figs.~\ref{fig2}(a), (b). Additionally, the stable dynamics was
monitored by plotting the $z'$ dependencies of the averaged
intensity $\int |\psi|^{4}d\mathbf{r}/N$ and mean-square radius
$\int r^{2}|\psi|^{2}d\mathbf{r}/N$. For stable propagation, these
quantities undergo small oscillations near the equilibrium values.
Note, that dipoles with sufficiently large (compared to
$\lambda_{th}$) values of $\lambda$ survive over huge distances
(many thousands of diffraction lengths) in the presence of quite
significant perturbations. We performed a series of runs for
$\lambda>200$ in the presence of strong initial noise. The initial
condition was taken in the form $\psi'(x',y')[1+\varepsilon
f(x',y')]$, where $\psi'(x',y')$ is the numerically calculated
exact dipole solution, $f(x',y')$ is the white gaussian noise with
variance $\sigma^{2}=1$ and the parameter of perturbation
$\varepsilon=0.1 \div 0.3$. Snapshots of $|\psi'(x',y')|$ at
different $z'$ for the case $\lambda=400$ and $\varepsilon=0.12$
are presented in Fig.~\ref{fig0}. One can see that the dipole
turns out to be extremely robust - even at $z'=2000$ one can not
detect any substantial distortion of the dipole shape. The
dipoles, however, become unstable (even if the initial noise is
very small) if $\lambda<\lambda_{th}$. The typical decay of the
unstable dipole near the threshold value of the rescaled
propagation constant is shown in Figs.~\ref{fig2}(c), (d).
 The situation, however, changes
below $\lambda_{cr}\approx 7.6$. Under this, the dipole splits in
two monopoles which move in the opposite directions without
changing their shape, i.~e. the monopoles just go away at
infinity. In Fig.~\ref{fig3} we plot the dipole energy $N_{dip}$
and the doubled energy $2N_{mon}$, where $N_{mon}$ is the energy
of the monopole soliton solution of Eq. (\ref{stat}), calculated
numerically, versus the propagation constant $\Lambda$. One can
see that the bound energy $\delta N=N_{dip}-2N_{mon}$ in the
dipole tends to almost zero as $\lambda$ approaches
$\lambda_{cr}=7.6$. This explains why the dipole with $\lambda\leq
\lambda_{cr}$ can be easily (i. e. under the action of extremely
small initial perturbations) split into two monopole-type
solitons.

The results of the numerical simulation can be illustrated through
the variational analysis. Equation (\ref{stat}) (in the scaled
variables) is the Euler-Lagrange equation for the Lagrangian
\begin{equation}
\label{L}
 \mathcal{L}=\int\left [|\nabla \psi'
|^{2}-\frac{1}{2}\theta\psi'^{2}+\lambda\psi'^{2}\right
]d^2\mathbf{r}.
\end{equation}
Taking a trial function in the form
\begin{equation}
\psi'=Ae^{-\beta^{2}[(x'-d/2)^{2}+y'^{2}]}-Ae^{-\beta^{2}[(x'+d/2)^{2}+y'^{2}]},
\end{equation}
where $A$, $\beta$ and $d$ are unknown parameters to be determined
by the variational procedure, and substituting it into Eq.
(\ref{L}), we get
\begin{equation}
\label{L1}
 \mathcal{L}=\pi A^{2}\left(2c_{1}-\frac{A^{2}}{4}c_{2}+\lambda
c_{3}\right),
\end{equation}
where
\begin{gather}
c_{1}=1-(1-\mu^{2}/2)e^{-\mu^{2}/2},\\
c_{2}=\frac{1}{\beta^{2}(1+\beta^{2})}[1+2e^{-\mu^{2}}+e^{-\mu^{2}/(1+\beta^{2})}\nonumber
\\
-4e^{-(3+2\beta^{2})\mu^{2}/4(1+\beta^{2})}],\\
c_{3}=\frac{1}{\beta^{2}}\left(1-e^{-\mu^{2}/2}\right),
\end{gather}
and, instead of $d$, we have introduced the variational parameter
$\mu=\beta d$. The optimum $A$ satisfies the equation $\partial
\mathcal{L}/\partial A=0$ which yields
\begin{equation}
\label{A}
 A^{2}=\frac{4c_{1}+2\lambda c_{3}}{c_{2}}.
\end{equation}
\begin{figure}
\includegraphics[width=3.4in]{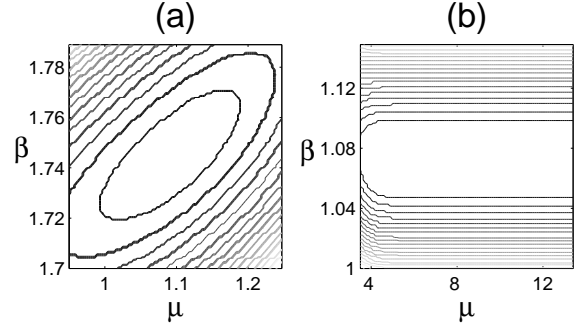}
\caption{\label{fig4} Contour maps of the function
$\mathcal{L}(\beta,\mu)$ : (a) $\lambda=40$; (b) $\lambda=7.6$}
\end{figure}
The Lagrangian Eq. (\ref{L1}), where $A^{2}$ is defined by Eq.
(\ref{A}), depends only on two unknown variational parameters
$\beta$ and $\mu$ (or, equivalently, $\beta$ and $d$) and can be
easily analyzed numerically. The topography of the function
$\mathcal{L}(\beta,\mu)$ depends on the rescaled propagation
constant $\lambda=\Lambda/\alpha^{2}$. There is the only minimum
if $\lambda > \lambda_{cr}$, where $\lambda_{cr}$ is some critical
value.
 The contours (level lines) of the function
$\mathcal{L}(\beta,\mu)$ for $\lambda=40$ are shown (in the
vicinity of the minimum) in Fig.~\ref{fig4} (a). In this case the
minimum takes place at $\beta=1.74$ and $\mu=1.1$. and corresponds
to the dipole solution presented in Fig.~\ref{fig1}. The amplitude
of the approximate analytical solution $A$ calculated from Eq.
(\ref{A}) and the parameters $\beta$ and $\mu$ are in agreement
with corresponding values estimated from the exact numerical
solution - the comparison of the variational analysis and the
direct numerical simulation is presented in Fig.~\ref{fig5} (a).
\begin{figure}
\includegraphics[width=3.4in]{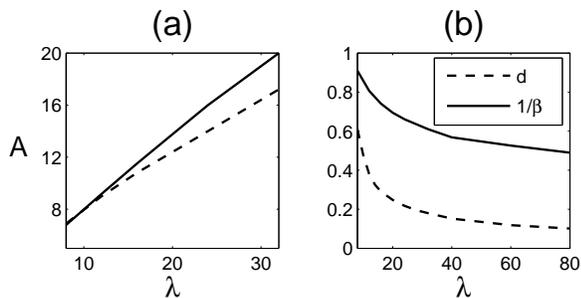}
\caption{\label{fig5} (a) The dependence of the dipole amplitude
$A$ on $\lambda$): solid curve - variational analysis; dashed
curve - direct numerical simulation. (b) Dipole distance $d$ and
the width $\beta^{-1}$ v. s. $\lambda$ (variational analysis)}
\end{figure}
The dependence of the width of the monopoles $\beta^{-1}$ in the
dipole and the "distance" $d$ between them on $\lambda$ is shown
in Fig.~\ref{fig5} (b). The topography of the function
$\mathcal{L}(\beta,\mu)$ in the vicinity of the minimum represents
a long narrow valley oriented at some angle to the $\mu$-axis.
Under this, the depth of the valley and the angle to the
$\mu$-axis decreases with decreasing $\lambda$. A similar
situation holds for all $\lambda
> \lambda_{cr}$ and we found $\lambda_{cr}\approx 7.6$.
The picture changes sharply at $\lambda=\lambda_{cr}$. The local
minimum disappears, the saddle point arises, and this corresponds
to the unstable dipole decaying into two monopoles. The contours
of the function $\mathcal{L}(\beta,\mu)$ in this case (for
$\lambda=7.6$) are shown in Fig.~\ref{fig4} (b). The found
critical value $\lambda_{cr}=7.6$ is in perfect agreement with the
results of direct numerical simulation (see Fig.~\ref{fig3}(b)).

\section{Summary and Conclusions}
\label{sec5}
We have investigated the main properties and stability of the
stationary two-dimensional localized solitary structures in the
nonlocal nonlinear media. We have studied both fundamental and
higher-order solitons; one-charge and multi-charge vortex solitons
with nonzero angular momentum; dipolar multisolitons. While the
fundamental soliton is always stable, the vortex solitons possess
a strong azimuthal instability which is eliminated only in the
strongly nonlocal regime. We have performed the linear stability
analysis and direct numerical simulations to investigate the
stability of vortices with arbitrary topological charge. We have
found the edge of the modulational instability and predicted the
threshold for the beam power of the robust vortex soliton. We
prove that in contrast to the nonlocal media with thermal
nonlinearity, the nonlinear response with the Gaussian-type kernel
can sustain not only single-charge but also multi-charge vortices.
We have investigated nonlocal  higher-order nonspinning solitons
which are the structures with the intensity distribution in the
form of a bright spot surrounded by the bright rings. We
theoretically predict an existence of stabilized higher-order
nonspining solitons in the nonlocal media. Finally, we have found
stationary dipole-like multisolitons which are the bound states of
the out-of-phase fundamental solitons. We have simulated
numerically the dynamics of the multisolitons in the presence of
initial noise and performed simple variational analysis. It turns
out that multisolitons are extremely robust at sufficiently high
input power in a highly nonlocal media. Therefore, these
predictions open the prospects for the experimental observations
of a wide class of stable coherent structures in various nonlocal
nonlinear media.

\begin{acknowledgments}
We are grateful to Yu.A. Zaliznyak for discussions and comments
about this paper. \end{acknowledgments}

\end{document}